\documentclass[12pt]{article}
\usepackage[T1]{fontenc}
\usepackage[utf8]{inputenc}
\usepackage[margin=1in]{geometry}
\usepackage[affil-it]{authblk} 
\usepackage{etoolbox}
\usepackage{lmodern}
\usepackage{cite}
\usepackage{changes}
\usepackage{microtype}
\usepackage{booktabs}
\usepackage{amssymb}
\usepackage{float}
\usepackage{textcomp}
\DeclareUnicodeCharacter{030C}{\v{c}}

\title{Triangular Quantum Photonic Devices with Integrated Detectors in Silicon Carbide}

\def\correspondingauthor{\footnote{Corresponding author: smajety@ucdavis.edu}}

\author[1] {Sridhar Majety \correspondingauthor{}}
\author[2] {Stefan Strohauer}
\author[1] {Pranta Saha}
\author[3] {Fabian Wietschorke}
\author[2] {Jonathan J. Finley}
\author[3] {Kai M\"uller}
\author[1] {Marina Radulaski}

\affil[1]{Department of Electrical and Computer Engineering, University of California, Davis, CA 95616, USA}
\affil[2]{Walter Schottky Institut, School of Natural Sciences, and MCQST, Technische Universit\"at M\"unchen, Am Coulombwall 4, 85748 Garching, Germany}
\affil[3]{Walter Schottky Institut, School of Computation, Information and Technology, and MCQST, Technische Universit\"at M\"unchen, Am Coulombwall 4, 85748 Garching, Germany}

\date{\vspace{-2em}}

\usepackage{graphicx}
\usepackage{amsmath}
\usepackage{braket}
\usepackage{titlesec}
\usepackage{siunitx}[=v2]
\usepackage[version=4]{mhchem}
\usepackage{multirow, makecell, caption}
\begin{document}

\maketitle
\vspace{-0.8cm}
\begin{abstract}

Triangular cross-section SiC photonic devices have been studied as an efficient and scalable route for integration of color centers into quantum hardware. In this work, we explore efficient collection and detection of color center emission in a triangular cross-section SiC waveguide by introducing a photonic crystal mirror on its one side and a superconducting nanowire single photon detector (SNSPD) on the other. 
Our modeled triangular cross-section devices with a randomly positioned emitter have a maximum coupling efficiency of \SI{89}{\percent} into the desired optical mode and a high coupling efficiency (> \SI{75}{\percent}) in more than half of the configurations. 
For the first time, \ce{NbTiN} thin films were sputtered on 4H-\ce{SiC} and the electrical and optical properties of the thin films were measured.
We found that the transport properties are similar to the case of \ce{NbTiN} on \ce{SiO2} substrates, while the extinction coefficient is up to \SI{50}{\percent} higher for \SI{1680}{\nm} wavelength.
Finally, we performed Finite-Difference Time-Domain simulations of triangular cross-section waveguide integrated with an SNSPD to identify optimal nanowire geometries for efficient detection of light from TE and TM polarized modes.

\end{abstract}

\section{Introduction}

Color centers in silicon carbide (SiC) have been a prominently studied quantum platform for applications in quantum information processing (QIP). 
Their unique combination of spectral homogeneity, emission at near infra-red and telecommunication wavelengths, long spin coherence, and spin-photon entangling processes offer key applications in quantum communication, simulation, computing, and sensing \cite{norman2021novel, son2020developing, castelletto2022silicon, zhang2020material, bathen2021manipulating, lukin2020integrated, majety2022quantum}. 

The optical response of a color center has a sharp peak called the zero phonon line (ZPL) and a broad phonon side band (PSB). 
Emission through ZPL, when there is no dephasing, results in indistinguishable photon creation necessary for QIP applications. 
Integration of color centers with photonic devices is an important step toward their performance improvement and scalability, as shown in explorations with SiC pillars \cite{radulaski2017scalable, parker2021infrared}, waveguides \cite{babin2022fabrication, lukin20204h}, and micro- and nano-resonators \cite{lukin20204h, crook2020purcell, bracher2017selective, calusine2014silicon, calusine2016cavity}. 
Moreover, recent results on growth and patterning of superconducting NbN on 3C-SiC \cite{martini2019single, martini2020electro} for use as the superconducting nanowire single photon detectors (SNSPDs) indicates that quantum circuitry can be further miniaturized and optimized.

The challenges in maintaining the pristine quality of color centers have led photonic integration away from the established nanofabrication processes and toward alternative approaches that require non-standard sample preparation \cite{ferro20153c, bracher2017selective, lukin20204h} or angled etching \cite{burek2012free, song2018high, babin2022fabrication}. 
The latter, in particular, has been explored for its potential in wafer-scale processing \cite{atikian2017freestanding}. 
Angle etching produces triangularly shaped devices whose photonic modes have been recently studied in terms of waveguide propagation \cite{majety2021quantum, babin2022fabrication}, photonic crystal band gap formation \cite{saha2022photonic}, photonic crystal cavity resonances \cite{majety2021quantum}, and grating coupler performance \cite{hadden2022design}.

\begin{figure}[!ht]
    \centering
    \includegraphics[scale=0.8]{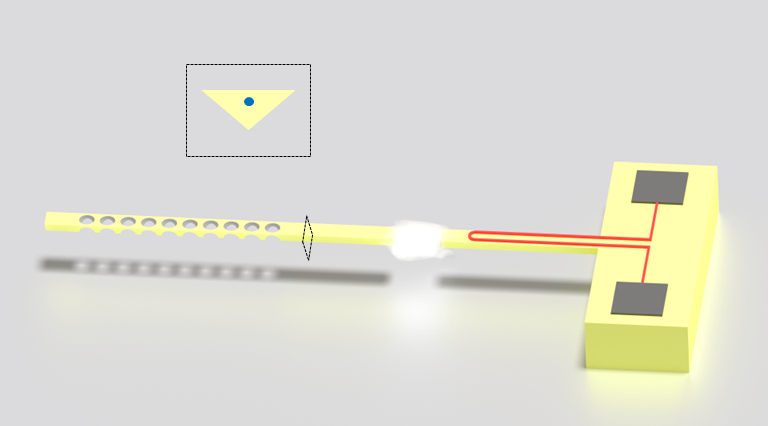}
    \caption{Illustration of SNSPD integrated onto a photonic device with a color center. Single color center is positioned in the waveguide region such that the dipole emission reflected by the photonic crystal mirror constructively interferes with the dipole emission into the waveguide region. The cloud region in the waveguide represents the segment where the pump light is filtered (e.g. by a ring oscillator or an inversely designed structure) to prevent any SNSPD saturation. Single photon detection is performed by the SNSPDs through absorption of the dipole emission in the waveguide. (Inset) The cross-section view of the structure, showing the color center (blue) optimally positioned at the centroid of the triangular waveguide profile.
    }
    \label{fig:fig1}
\end{figure}

In this paper, we expand on the studies of triangular SiC photonics by focusing on methods to optimize collection and detection of color center light by integration with a photonic crystal mirror (PCM) and an SNSPD, as illustrated in Figure \ref{fig:fig1}. 
In particular, we study the case of the nitrogen vacancy (NV) in 4H-SiC with emission wavelengths in the range \SIrange{1176}{1243}{\nm} \cite{mu2020coherent, sato2019formation, wang2020coherent}. 
When placed in a symmetric waveguide, at least half of the light is lost in an undesired direction, but this can be circumvented by the addition of a PCM. 
PCM has a periodic variation of refractive index, which results in the formation of photonic band gaps. Understanding the formation of these band gaps is necessary for applications in selective reflection of the ZPL and PSB emissions with the goal of enhancing light collection efficiency \cite{saha2022photonic}. 
The other side of the waveguide deals with the efficient detection through hybrid integration with NbTiN SNSPDs. The SNSPDs are the leading detectors in terms of quantum efficiency, low dark counts and low timing jitter. 
Moreover, they offer an on-chip detection that overcomes the losses involved in the off-chip detection, thus boosting the success rate of QIP protocols \cite{najafi2015chip}.
However, depending on the type of excitation, the pump laser used for an off-resonant excitation of the color center can scatter into the detector region and saturate the SNSPD. There have been successful demonstrations to suppress the stray pump laser by coating the backside of the chip with an absorbing material \cite{reithmaier2015chip} and masking the inactive regions of the chip \cite{schwartz2018fully}. To further suppress the pump laser coupled into the waveguide, spectral filters like ring resonators \cite{elshaari2017chip, kim2020hybrid} and inverse-designed structures \cite{molesky2018inverse, dory2019inverse} can be integrated in the waveguide region before the detectors, with implementation in the illustrated cloud segment in Figure \ref{fig:fig1}.

We first find the 1D photonic crystal (PC) conditions for band gap formation in triangular geometry on the desired wavelength domain (NV in 4H-SiC in this work) using the Plane Wave Expansion (PWE) and the Finite-Difference Time-Domain (FDTD) methods. 
We then utilize the optimal PC design as a PCM in a triangular waveguide and study the role of color center positioning in obtaining high emission coupling to the propagating optical mode. 
Finally, in a combination of experimental and modeling approaches, we study the triangular waveguide mode absorption by a NbTiN SNSPD.

\section{Photonic crystals for efficient propagation of color center emission in a triangular waveguide}

Color centers have an optical dipole-like emission covering a solid angle of 4$\pi$ \cite{babin2022fabrication}. 
This causes \SI{50}{\percent} higher than in spin-entangled photon loss in the triangular cross-section waveguide resulting in low collection efficiency. 
An \textit{in situ} mirror, capable of reflecting color center emission, should theoretically double the collection efficiency. 
It is possible to design such a structure artificially by periodic variation of the dielectric contrast, known as the photonic crystal \cite{joannopoulos1997photonic}. 
For enhanced collection, we deliberately open up a gap possessing the NV center emission wavelength in the photonic bands, which is achievable by making a particular choice of PC parameters \cite{johnson1999guided}.

Waveguides support light propagation in a particular direction via total internal reflection. 
We implement 1D PC in our triangular cross-section waveguide to introduce asymmetry for redirecting the color center emission. 
However, 1D PCMs have mostly been realized in conventional rectangular geometry \cite{notomi2008ultrahigh,quan2011deterministic,huang2018elliptical}, and photonics with triangular geometry has been focused on making efficient active photonic devices \cite{burek2014triangulardiamond,majety2021quantum,majety2022quantum}. 
Photonic band structures in triangular geometry have not been studied in detail. 
In this light, our recent work provides an insightful explanation of the dispersion relations in SiC triangular geometry \cite{saha2022photonic}. 
From the band structures obtained from Plane Wave Expansion (PWE) method, we use Finite-Difference Time Domain (FDTD) method to analyze the performance of the triangular-cross section 1D PCM and the dependence of the coupling efficiency to the waveguide mode on the emitter position with respect to the PCM. 

\subsection{Considerations for integrating color centers into triangular cross-section devices}
Color centers in SiC have multiple orientations due to the presence of inequivalent lattice sites (h, k). 
The dipole orientations of the color centers can be parallel (silicon vacancy in SiC) or at an angle (NV center in SiC) to the crystal axis (c-axis). 
In a photonic device fabricated from commonly available 4H-SiC, grown along the c-plane, this would result in emissions strongly coupling to the fundamental TM (f-TM) mode \cite{radulaski2017scalable}. 
On the other hand, it has been shown that using 4H-SiC grown along the a-plane, the color center emission can be predominantly coupled to the fundamental TE (f-TE) mode \cite{babin2022fabrication}.

Triangular cross-section waveguides support f-TE/TM modes and other higher order modes \cite{majety2021quantum}. 
The fraction of a color center emission coupling to a supported waveguide mode depends on the position of the emitter within the triangular cross-section. 
Highest coupling to a mode can be achieved by positioning the emitter at the maximum intensity point of that mode. 
The maximum intensity point of the f-TE mode is located approximately at the centroid of the triangle, which is an optimal color center position for achieving highest coupling efficiency to the f-TE mode \cite{majety2021quantum, babin2022fabrication}. 

The depth of the centroid from the surface depends on the width and etch angle of the triangular cross-section waveguide. 
These dimensions can be decided on the basis of the implantation capabilities, to ensure that the implanted color centers achieve the highest coupling efficiency to the f-TE mode. 
For a given emission wavelength, a triangular cross-section waveguide has an optimal width at which it has single mode propagation (f-TE), with high (> \SI{80}{\percent}) coupling to the f-TE mode \cite{babin2022fabrication}. 
Single mode propagation in the waveguide is key for applications in QIP \cite{yusof2019ribwvg,westig2020singlemode}. 
The optimal width of the triangular cross-section devices increases with the etch angle and emission wavelength. 
Hence, triangular cross-section devices with larger etch angles have more fabrication friendly dimensions. Moreover, they require shallow implantation depths, which causes less damage to the crystal and having an emitter closer to the surface is useful for SNSPD absorption. 
Advantages like single mode propagation with high coupling efficiency, less damage to the crystal and better detection in SNSPD outweighs the overall effect of the color center emission lost into the f-TM or other modes.

\subsection{Waveguide design with TE and TM band gap}
\label{PWE}
NV center emission in SiC has both TE- and TM-like components \cite{wang2020NVSiC}. Therefore, a PCM with polarization-independence is necessary to increase collection efficiency. 
For broadband reflectivity, such a mirror must have as large a complete photonic band gap as possible \cite{zhao2011broadmirror, wang2001effects}. 
Single mode propagation \cite{yusof2019ribwvg,westig2020singlemode} and positioning of emitter at the centroid of the triangular cross-section \cite{babin2022fabrication, majety2021quantum} ensures that most of the color center emission is coupled into the waveguide. 
It was shown that \SI{45}{\degree} etch angle has predicted the best results for the complete band gap \cite{saha2022photonic}. 
But in this study, an etch angle of \SI{60}{\degree} was chosen because the optimal emitter position is closer to the top surface, offering realistic implantation depths (\SI{70}{\nm}) and efficient absorption in the SNSPDs. 
Moreover, a \SI{60}{\degree} etch angle supports single mode propagation for larger dimensions than in the case of \SI{45}{\degree} \cite{babin2022fabrication}, resulting in a more fabrication friendly waveguide design. 
The width of the waveguide was set to 800 nm, as it supports the f-TE and f-TM modes only.   

\begin{figure}[htp]
    \centering
    \includegraphics[width = \textwidth]{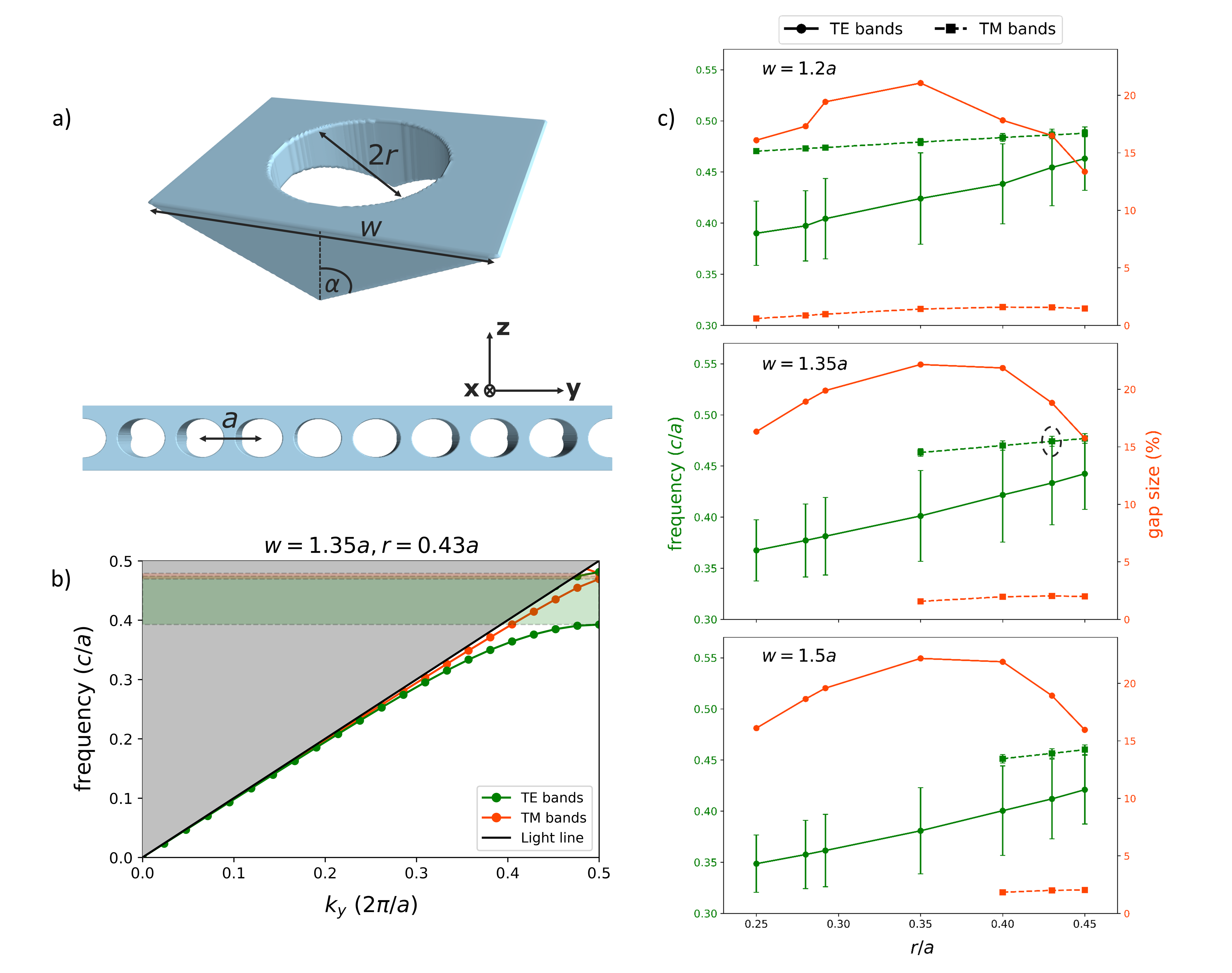}
    \caption{a) Unit cell of the 1D photonic crystal structure. b) Band structure for $\alpha$ = \SI{60}{\degree}, $w = 1.35a$, and $r = 0.43a$. c) TE/TM photonic band gap and gap size as a function of normalized hole radii for $w=1.2a, 1.35a, 1.5a$ respectively. Center point indicates midgap frequency and error bar indicates gap width for the corresponding TE/TM modes. The dashed circle shows the TE-TM gap overlap for the band structure shown in b).
    }
    \label{fig:fig2}
\end{figure}

The 1D PC is made of cylindrical air holes in a triangular cross-section SiC waveguide as shown in Figure \ref{fig:fig2}a. 
The parameters are as follows: periodicity along the $y$ axis $a$, waveguide width $w$, hole radius $r$, and etch angle $\alpha$, which corresponds to half-angle at the apex. 
The refractive index of SiC is considered 2.6 for all the simulations. In this study, we focused on the dispersion relations for $\alpha$ = \SI{60}{\degree} w.r.t $w$ and $r$. We vary $w$ from $1.2a$ to $2.25a$ and $r$ from $0.25a$ to $0.45a$. 
The band structure in Figure \ref{fig:fig2}b exhibits three properties: i) a wide TE band gap, ii) a narrow TM band gap, and iii) a partial overlap between TE and TM band gaps, which is the complete photonic band gap. 
Figure \ref{fig:fig2}c delineates the changes in the TE/TM gap $(c/a)$ and the corresponding gap size (gap-midgap ratio) with $r/a$ for $w$ values of 1.2a, 1.35a and 1.5a. 
Midgap frequency increases with $r/a$ for both TE and TM band gaps as volumetric fill factor (VFF) and effective refractive index ($n_\mathrm{eff}$) have an inverse relationship with $r/a$ \cite{saha2022photonic}. 
Gap size initially increases with $r/a$ as fewer propagating modes are supported, but declines after the maximum value is reached. 
Though this trend is observed for both TE and TM bands, the TM band gap completely vanishes for larger widths ($w > 1.5a$). 
Moreover, the highest TE gap size ($\sim$\SI{22}{\percent}) is an order magnitude higher compared to the highest TM gap size ($\sim$\SI{2}{\percent}), which is attributed to the connectivity between high-dielectric regions \cite{joannopoulos1997photonic}. 
In general, complete band gap mostly occurs for smaller widths with larger hole radii as depicted in Figure \ref{fig:fig2}c. 
Nevertheless, the choice of parameters for designing the reflector is made in such a way that there is a balance between optimal band gap size and reproducibility of the nanofabrication.  

\subsection{SiC waveguide design with a PCM}
In this section, we use the PC parameters from the PWE method and simulate a triangular cross-section waveguide with a reflector, using the Finite-Difference Time-Domain (FDTD) package in Lumerical \cite{majety2021quantum, babin2022fabrication}. 
This allows for a comparison of the band gap position estimated from the two different methods, PWE and FDTD. 
The FDTD method uses the Fourier transform of the time-domain signal to record the field data in the frequency domain. 
To increase the collection efficiency of the photons from the ZPL and PSB of the NV center in SiC, the PC parameters were chosen to have a band gap spanning these emission wavelengths. 
As mentioned in Section \ref{PWE}, the size of the TM band gap (and thus the complete band gap) is significantly smaller than the size of the TE band gap, limiting the reflected wavelengths of both TE- and TM-like components of the NV center emission.

In this study, we focus on two sets of parameters for the PC 1) complete band gap and 2) NV optimized band gap, for a waveguide with $w$ = \SI{800}{\nm} and $\alpha$ = \SI{60}{\degree}. 
In the first scenario, the complete band gap was achieved using PC parameters of $r$ = \SI{237}{\nm} and $a$ = \SI{593}{\nm}, with a complete band gap from \SIrange{1268}{1274}{\nm} (TE band gap: \SIrange{1268}{1579}{\nm}, TM band gap: \SIrange{1249}{1274}{\nm}), which is outside the emission wavelength range of the NV center. 
In the second scenario, PC parameters of $r$ = \SI{213}{\nm} and $a$ = \SI{533}{\nm} were chosen, resulting in both TE (\SIrange{1200}{1494}{\nm}) and TM (\SIrange{1170}{1192}{\nm}) band gaps individually overlapping with the emission wavelength range of NV centers, but no complete band gap. 

\begin{figure}[!ht]
    \centering
    \includegraphics[width=\textwidth]{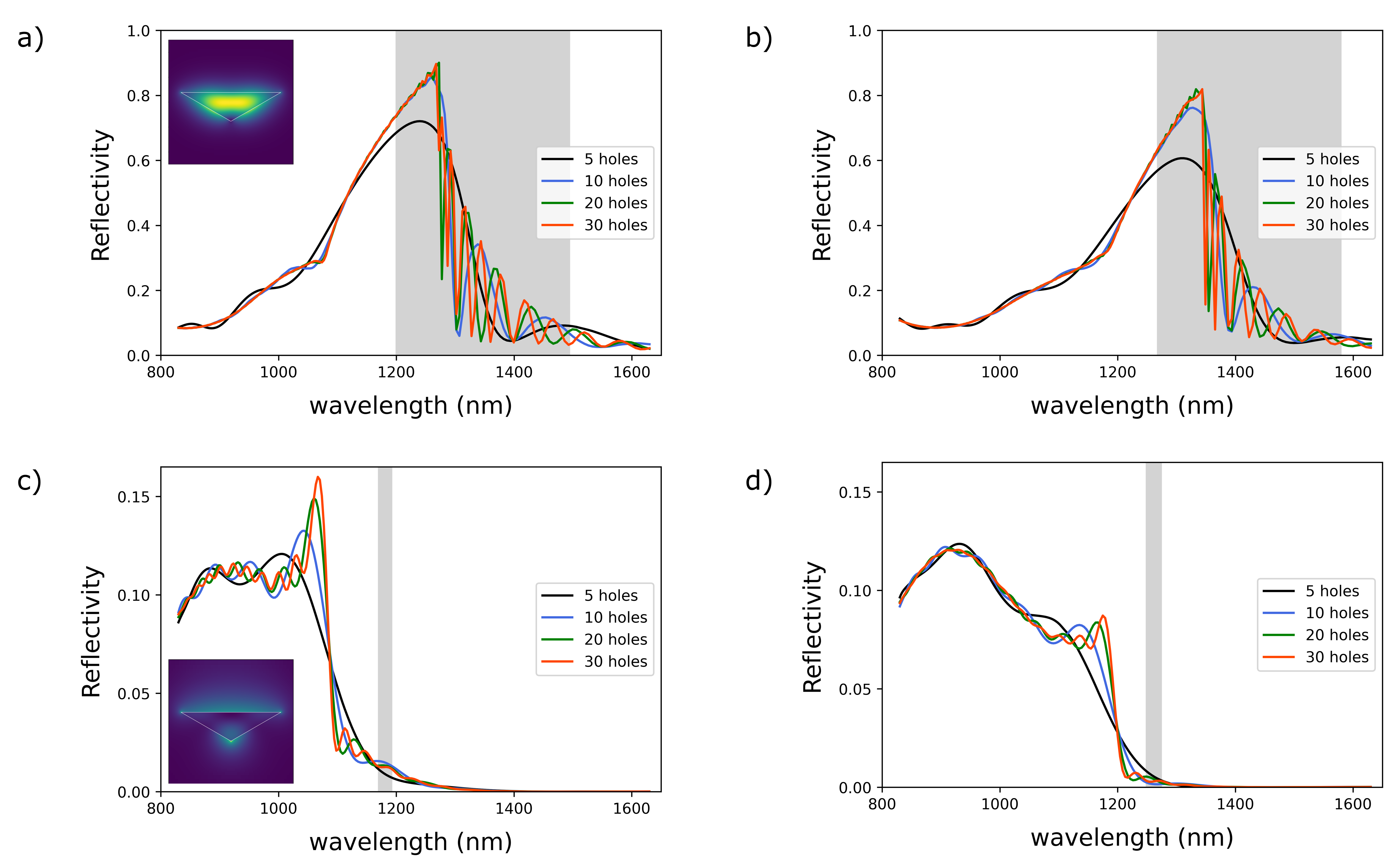}
    \caption{The reflectivity of photonic crystal mirror as a function of wavelength for varying number of air holes. Grey regions indicate the band gap values calculated from the PWE method. a) Reflectivity of the fundamental TE mode for a NV optimized design. Inset shows the mode profile of the fundamental TE mode of the waveguide at \SI{1230}{\nm}. b) Reflectivity of the fundamental TE mode for a complete band gap design. c) Reflectivity of the fundamental TM mode for a NV optimized design. Inset shows the mode profile of the fundamental TM mode of the waveguide at \SI{1230}{\nm}. d) Reflectivity of the fundamental TM mode for a complete band gap design.
    }
    \label{fig:fig3}
\end{figure}

As the number of air holes increases, the maximum reflectivity of the PCM increases and saturates for 20 air holes, as shown in Figure \ref{fig:fig3}a-b. 
It is observed that the FDTD calculated wavelength for maximum reflectivity deviates from the center of the PWE calculated bandgap (grey region) \cite{aravantinos2014phononic}. 
The reflectivity curve has a full width half maximum of $\sim$\SI{200}{\nm}, large enough to reflect the ZPL and the entire PSB of the color center emission. 
At \SI{1230}{\nm}, the NV optimized design has $\sim$\SI{85}{\percent} reflectivity for the f-TE mode. 
The reflectivity at a given wavelength depends on two factors: 1) the effective index of the f-TE mode ($n_{\textnormal{eff}}$) and 2) the fraction of color center emission coupled into the waveguide. While $n_{\textnormal{eff}}$ decreases monotonically with an increased wavelength, the coupling efficiency rises to the optimal point (designed at 1230 nm) and subsequently decreases. The combination of the two effects cause the reflectivity to gradually increase for the short wavelengths and sharply drop for the long wavelengths.
In the case of the f-TM modes, the reflectivity values are relatively lower due to the low coupling efficiency to the f-TM mode ($\sim$\SI{20}{\percent}) and evanescent loss to the surroundings ($n_{\textnormal{eff}}$ = 1.09).

\begin{figure}[!ht]
    \centering
    \includegraphics[width=0.75\textwidth]{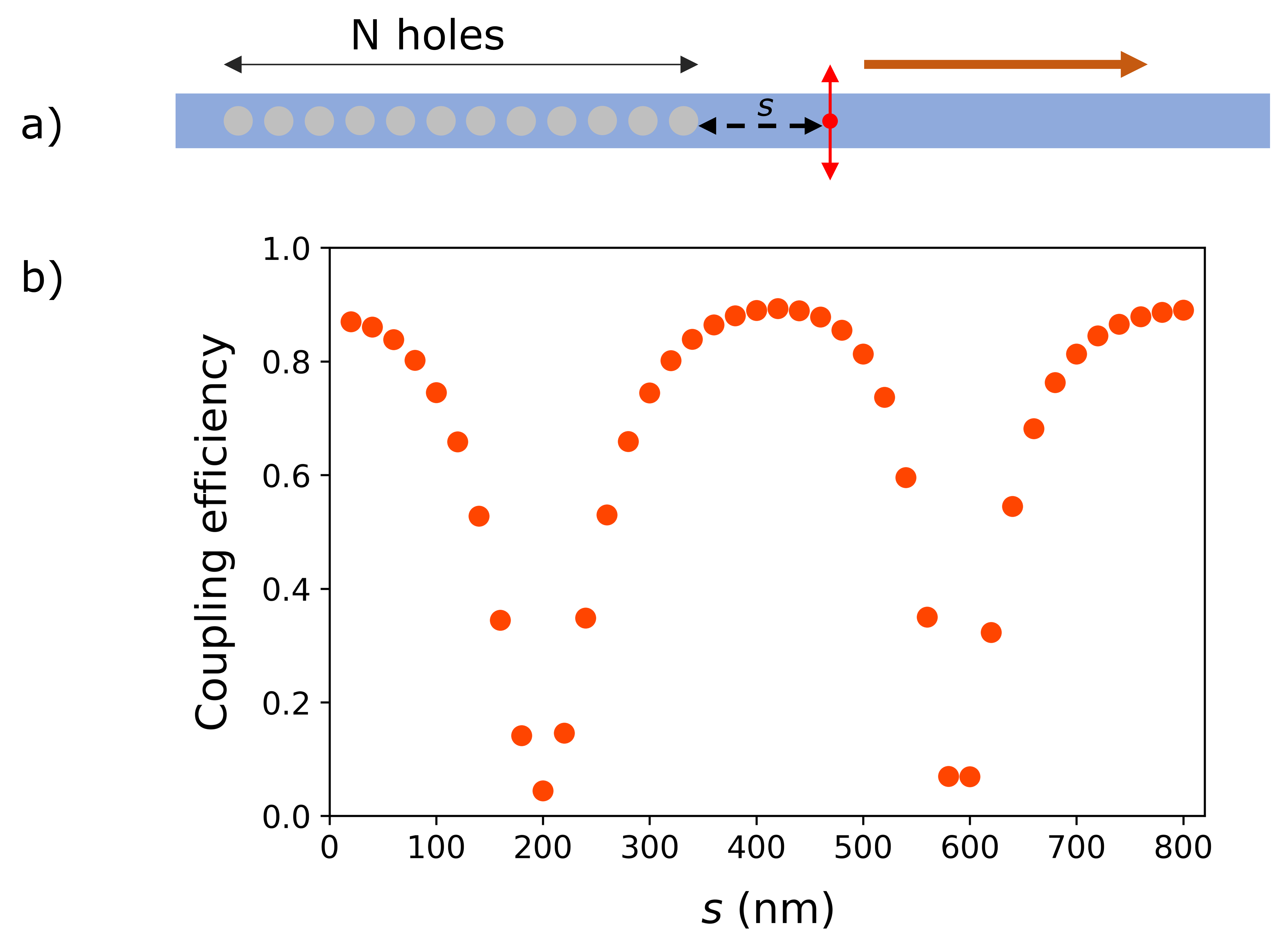}
    \caption{a) Top view of the waveguide design with reflector having N air holes. The color center (red dot with red arrows indicating the dipole orientation) is positioned at a distance $s$ from the edge of the first hole and its emission propagation indicated by the arrow (brown). b) The variation of the coupling efficiency of the color center emission to the fundamental TE mode, in the direction of propagation, as a function of the color center's position ($s$) from the photonic crystal. 
    }
    \label{fig:fig4}
\end{figure}

When a dipole emitter (\SI{1230}{\nm}) is positioned at the centroid of the triangular cross-section ($\sim$\SI{70}{\nm} depth) of the waveguide ($w$ = \SI{800}{\nm}, $\alpha$ = \SI{60}{\degree}), \SI{44}{\percent} of the \SI{50}{\percent} light in each direction of the waveguide is coupled to the f-TE mode ($n_{\textnormal{eff}}$ = 1.574). 
In the presence of a PCM, the \SI{50}{\percent} dipole emission directed toward the PCM is reflected back towards the emitter, causing interference with the emission propagating in the waveguide region (indicated by a brown arrow in Figure \ref{fig:fig4}a). 
We observe that as the distance between the emitter and the PCM ($s$) is varied, the total light coupled to the f-TE mode varies periodically, as shown in Figure \ref{fig:fig4}b. 
Constructive (destructive) interference occurs if the additional distance 2$s$ traveled by the reflected wave results in a phase difference that is an even (odd) multiple of $\pi$. 
The periodicity of the coupling efficiency matches with the calculated value ($\lambda/n_{\textnormal{eff}}$ = \SI{390}{\nm}) and the maximum coupling efficiency of \SI{89}{\percent} is achieved for values of $s$ that are integer multiples of $\sim$\SI{390}{\nm}. 
There exists a \SI{200}{\nm} region with coupling efficiency >\SI{75}{\percent}. 
Thus, any randomly positioned emitter in the waveguide has at least \SI{50}{\percent} chance of being at a position with high coupling efficiency.

\section{Integrated SNSPDs for efficient detection of light in triangular waveguides}
For single-photon detection, SNSPDs \cite{GolTsman2001} provide unrivaled performance metrics such as high detection efficiency, including at \SI{1550}{\nm} wavelength \cite{Korzh2020}, high count rate, low dark-count rate \cite{Shibata2015}, and high temporal resolution \cite{Reddy2020}. 
As such, they have found many applications in quantum communication and information processing as well as in high energy physics for charged particle detection \cite{EsmaeilZadeh2021, You2020, Polakovic2020}.
Compared to the top-illumination of meander SNSPDs and the enhancement of their detection efficiency by use of an optical cavity with a mirror underneath \cite{Marsili2013, Redaelli2016}, short waveguide integrated SNSPDs avoid coupling losses when being interfaced with on-chip devices and provide smaller recovery times as a result of their lower kinetic inductance \cite{Reithmaier2013, Ferrari2018}.
Due to their superior performance metrics, in this section we study the potential of integrating SNSPDs onto triangular cross-section \ce{SiC} waveguides to detect single-photon emission from color centers embedded in the waveguide.
We start by investigating the fundamental properties of \ce{NbTiN} films on \ce{SiC} substrates and subsequently provide performance estimates for SNSPDs on \ce{SiC}.
Next, we use FDTD simulations to determine the coupling efficiency from triangular \ce{SiC} waveguides to SNSPDs integrated on top of the waveguide and its dependence on the detector geometry.

\subsection{Optical and transport properties of \ce{NbTiN} thin films on \ce{SiC}}
\label{SNSPD-exp}
In this section, we investigate the optical and the transport properties of \ce{NbTiN} thin films on \ce{SiC} substrates.
For this, we used DC reactive magnetron sputtering at room temperature with an \ce{Ar}/\ce{N2} atmosphere and a sputtering target with the stoichiometry \ce{Nb_{0.7}Ti_{0.3}} to fabricate the \ce{NbTiN} films studied in the following.

\begin{figure}[htp]
    \centering
    \includegraphics[width = 0.8\textwidth]{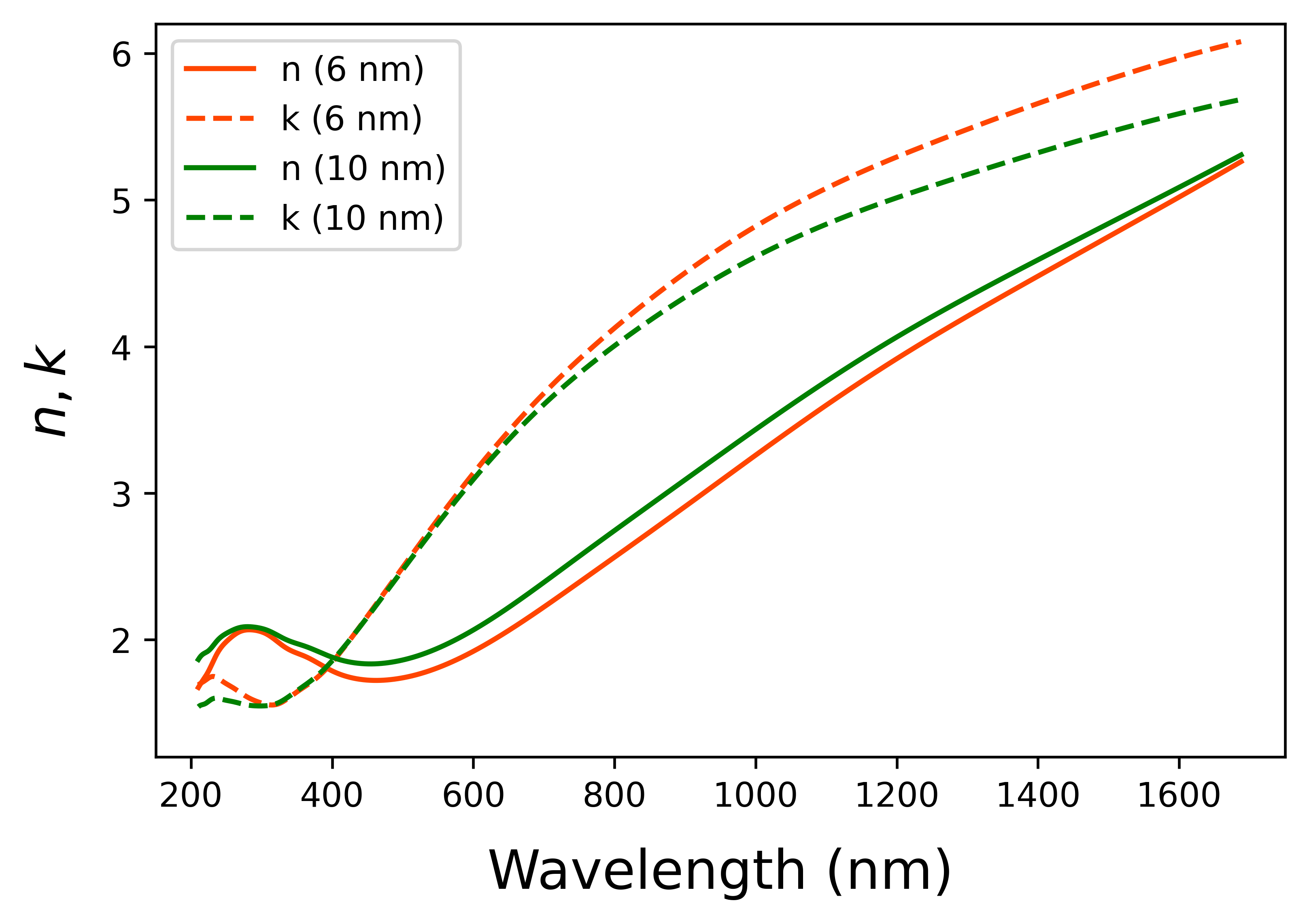}
    \caption{Extinction coefficient $k$ and refractive index $n$ of \SI{6}{\nm} and \SI{10}{\nm} \ce{NbTiN} films on 4H-\ce{SiC} substrates.
    }
    \label{fig:fig5}
\end{figure}

For simulation of absorption in NbTiN SNSPDs on triangular cross-section SiC waveguides, the optical constants of the deposited superconductor are essential, especially because the optical and the transport properties vary with deposition parameters and the underlying substrates \cite{Makise2011}.
We used a variable angle spectroscopic ellipsometer, Model M-2000 of J.A. Woollam Co., to measure the optical constants over a wavelength range of \SIrange{210}{1687}{\nm}.
As \ce{NbTiN} is absorbing at all measurement wavelengths and one cannot resort to a transparent wavelength window, it is difficult to decouple and determine the optical constants and film thickness simultaneously due to correlations between both. 
However, this correlation can be overcome by measuring with the interference enhancement method, where the film is measured on an additional thick transparent dielectric layer and at multiple angles  \cite{Banerjee2018, Hilfiker2008, Hilfiker2008a}.
As we measured the \ce{NbTiN} films on two-sided polished \ce{SiC} substrates, two additional challenges emerge: 
The occurrence of reflections on the back side of the substrate, as well as the fact that \ce{SiC} is a birefringent material.
However, similar to the interference enhancement method, the anisotropy helps in decoupling thickness and optical constants for the \ce{NbTiN} film \cite{Priv_Comm_ThomasWagner_anisotropyInterferenceEnhancement}.
Including backside reflections of the uniaxial 4H-\ce{SiC} substrate in the model, we obtain a fitting of the measured data with a mean square error less than 5, indicating good agreement between the model and the measured data. 
The resulting optical constants, shown in Fig.~\ref{fig:fig5}, reveal a smaller extinction coefficient $k$ and a higher refractive index $n$ for the thicker film and provide the basis for the subsequent FDTD simulations.
Compared to the optical constants of a \SI{5.5}{\nm} thick \ce{NbTiN} film grown by sputtering on silicon substrates in the publication of Banerjee \cite{Banerjee2018}, we obtain similar results, although our data shows no features in the range from \SIrange{500}{1000}{\nm} and our refractive index is \textasciitilde\SI{20}{\percent} and the extinction coefficient \textasciitilde\SI{50}{\percent} higher for a wavelength of \SI{1680}{\nm}.
This agrees with the fact that the growth and properties of polycrystalline materials such as \ce{NbTiN} depend on the underlying substrate.

\begin{figure}[!ht]
    \centering
    \includegraphics[width = \textwidth]{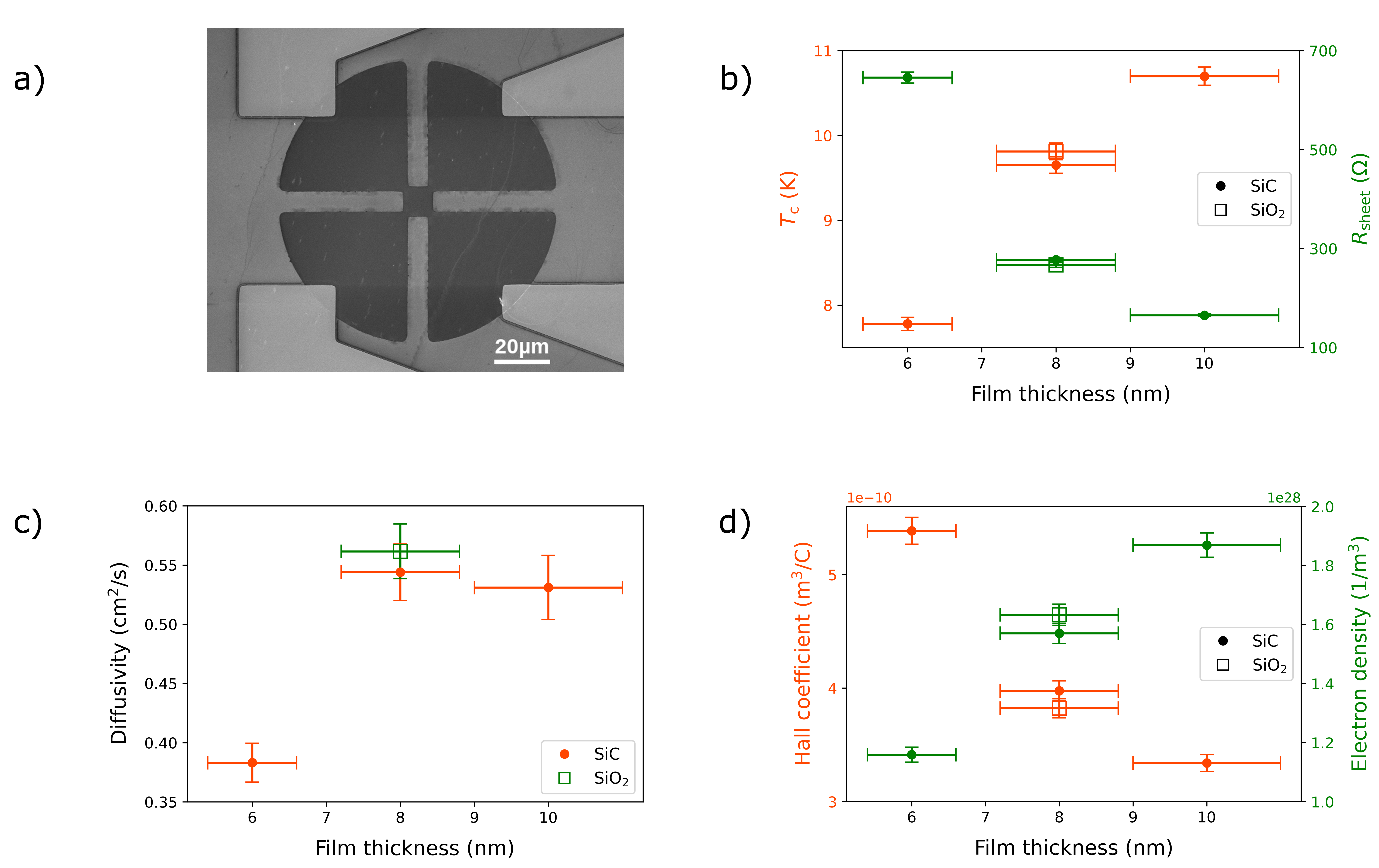}
    \caption{Critical temperature and transport measurements of sputtered NbTiN on 4H-SiC substrates. a) shows an SEM image of the fabricated cloverleaf structures that were used for performing the magneto-transport measurements. In graphs b), c), and d), the measured data is shown for films of nominal thicknesses \SI{6}{\nm},  \SI{8}{\nm}, and \SI{10}{\nm}.
    }
    \label{fig:fig6}
\end{figure}

As fabrication and characterization of SNSPDs is technologically challenging and time consuming, it is beneficial to estimate the detector performance directly from properties of the sputtered films---especially when it comes to new superconducting materials or different substrates.
To assess, for example, possible operating temperatures of the envisioned SNSPDs, it is essential to know the device's critical temperature $T_\mathrm{c}$ at which it switches from the superconducting to the normal conducting state.
As this temperature is reduced for nanowires compared to thin films or bulk material \cite{Holzman2019, Tanner2012}, the critical temperature of the film provides an upper limit for reachable SNSPD critical temperatures.
Furthermore, by measuring fundamental film transport properties such as sheet resistance $R_\mathrm{sheet}$, quasiparticle diffusivity $D$ \cite{Bartolf2016, Catelani2022}, and Hall coefficient $R_\mathrm{H}$, we can estimate key detector metrics such as depairing current $I_\mathrm{dep}$ (an upper limit for the detector's switching current) and fall time $\tau_\mathrm{fall}$, and obtain information about disorder in the film (via the deduced Ioffe-Regel parameter $k_\mathrm{F}l$ \cite{Hazra2018}).
The depairing current density at a temperature $T$ can be calculated according to the Bardeen model \cite{Korneeva2018, Clem2012, Bardeen1962, Bartolf2016} from the superconductor's energy gap $\Delta(\SI{0}{\K}) = 2.0\, k_\mathrm{B}\, T_\mathrm{c}$ with the factor $2.0$ for \ce{NbTiN} according to \cite{Khan2022, Lap2021}, the elementary charge $e$, the resistivity in the normal conducting state $\rho_\mathrm{nc}$, and the quasiparticle diffusivity $D$ by
\begin{equation}
j_\mathrm{dep}(T) = 0.74\cdot
    \frac{
    \left(\Delta(\SI{0}{\K})\right)^{3/2}
    }{
    e \rho_\mathrm{nc} \sqrt{\hbar D}
    }
    \left(1 - \left(\frac{T}{T_\mathrm{c}}\right)^2 \right)^{3/2}
    \;.
\end{equation}
A higher switching current of the detector also means a higher amplitude of the detection pulse, which is beneficial for the signal-to-noise ratio.

Moreover, we can calculate the kinetic inductance from the sheet resistance $R_\mathrm{sheet}$, the energy gap $\Delta$, and the nanowire width $w$ and length $l$ \cite{Bartolf2016, Schmidt1997}.
\begin{equation}
    L_\mathrm{k}
    = \frac{\hbar\, R_\mathrm{sheet}}{\pi\, \Delta(\SI{0}{\K})} \frac{l}{w}
\end{equation}
that can be used to estimate the fall time of a detection pulse \cite{Kerman2006, Smirnov2016} by
\begin{equation}
    \tau_\mathrm{fall}
    = \frac{L_\mathrm{k}}{R_\mathrm{load}}
    \;,
\end{equation}
with $R_\mathrm{load}$ denoting the load resistor, which can be either a shunt resistor or the readout electronics (typically \SI{50}{\ohm}).
From the fall time, we can also estimate the maximum count rate by $1/\tau_\mathrm{fall}$.
In Fig.~\ref{fig:fig6}, we show the measured transport properties of \SI{6}{\nm},  \SI{8}{\nm}, and \SI{10}{\nm} sputtered NbTiN\footnote{These are the nominal thicknesses set during the sputtering process and are used in conjunction with the measured deposition rate to calculate the necessary sputtering time. For \SI{6}{\nm} and \SI{10}{\nm}, the thicknesses have also been fitted within the optical model for the ellipsometry data and yielded \SI{5.48+-0.2}{\nm} and \SI{9.62+-0.26}{\nm}.} on 4H-SiC substrates with one \SI{8}{\nm} film of the same sputtering round on a silicon wafer with thermally grown \ce{SiO2} as reference substrate.
From this fundamental data, we calculate estimates for key detector metrics in Tab.~\ref{tab:calculated_SSPD_properties}.
Despite the difference in refractive index and extinction coefficient found above, a comparison of the measured and derived performance metrics for \ce{NbTiN} SNSPDs on \ce{SiC} substrates with reference to \ce{NbTiN} SNSPDs on \ce{SiO2} substrate yields similar values.
Also, our measured transport data agree with the literature \cite{Sidorova2021, Hazra2018, Steinhauer2020}.
However, it seems worthwhile to note that our data for the electron density is lower than the values found in the literature.
Nevertheless, when comparing the data in \cite{Hazra2018, Sidorova2021}, the electron density shows a dependence on the stoichiometry of \ce{Nb_{x}Ti_{1-x}N}, and our case with $x=0.7$ fits into the picture of having a minimum between $x=0.66$ and $x=0.86$.

We conclude that choosing \ce{SiC} substrates in combination with our sputtering recipe yields films that have higher absorption while maintaining similar transport properties.
Thus, geometrically identical detectors can be expected to have a higher detection efficiency with similar electrical performance on \ce{SiC} compared to \ce{SiO2} substrates.
At the same time this can be used to reduce the optical active area of a detector to enhance its electrical properties (for example reducing the fall time) while keeping the same detection efficiency.
It is worthwhile to note that the absolute values we obtained for the fall time of SNSPDs are higher than typically measured fall times that are in the order of \SI{10}{\ns} for \SI[parse-numbers=false]{10\times10}{\um} SNSPDs on \ce{SiO2} substrates (measured, e.g., in \cite{Kerman2006} as well as in our labs).
We speculate that the fabrication process of detectors modifies the transport properties of the film such that the kinetic inductivity and thus the fall time are increased compared to the film-derived ones.

\begin{table}
	\centering
	\caption{
	Estimated SNSPD performance metrics from measured film properties.
    Estimated SNSPD performance metrics from measured film properties.
    For the derived numbers of the meander SNSPDs we consider \SI[parse-numbers=false]{10\times10}{\um} detectors with \SI{100}{\nm} wide wires and a fill factor of \SI{50}{\percent}, while for the double loop SNSPD we assume a \SI{100}{\nm} wide and \SI{40}{\um} long nanowire, similar to the maximum wire length used in the simulations presented in Figures~\ref{fig:fig7} and \ref{fig:fig8}.
	}
	\label{tab:calculated_SSPD_properties}
\begin{tabular}{
c
S[table-number-alignment = center]
S[table-number-alignment = center]
S[table-number-alignment = center]
S[table-number-alignment = center]
S[table-number-alignment = center]
S[table-number-alignment = center]
S[table-number-alignment = center]
}
 \toprule
    & & & & \multicolumn{2}{c}{Meander SNSPD} & \multicolumn{2}{c}{Double loop SNSPD} \\
    \cmidrule(rl){5-6} \cmidrule(rl){7-8}
    {Substrate}  & {$d$/nm}   & {$I_\mathrm{dep}(\SI{0}{\K})$/\textmu A}  & {$I_\mathrm{dep}(\SI{4}{\K})$/\textmu A}    & {$L_\mathrm{k}$/\textmu H} & {$\tau_\mathrm{fall}$/ns} & {$L_\mathrm{k}$/\textmu H} & {$\tau_\mathrm{fall}$/ns}  \\
    \midrule
    \multirow{3}{*}{\ce{SiC}}                & 6   & 14.1 & 8.9  & 3.2  & 63.5 & 0.254  & 5.08     \\
                                        & 8   & 38.1 & 28.7 & 1.1  & 22.0 & 0.088  & 1.76    \\
                                        & 10  & 75.5 & 60.3 & 0.6  & 11.8 & 0.047  & 0.95    \\
    \midrule
    \ce{SiO2}                           & 8   & 40.0 & 30.4 & 1.0  & 20.8 & 0.083  & 1.66    \\
  \bottomrule
\end{tabular}
\end{table}

\subsection{Model of an integrated SNSPD with a triangular cross-section SiC waveguide}

\begin{figure}[!ht]
    \centering
    \includegraphics[width = \textwidth]{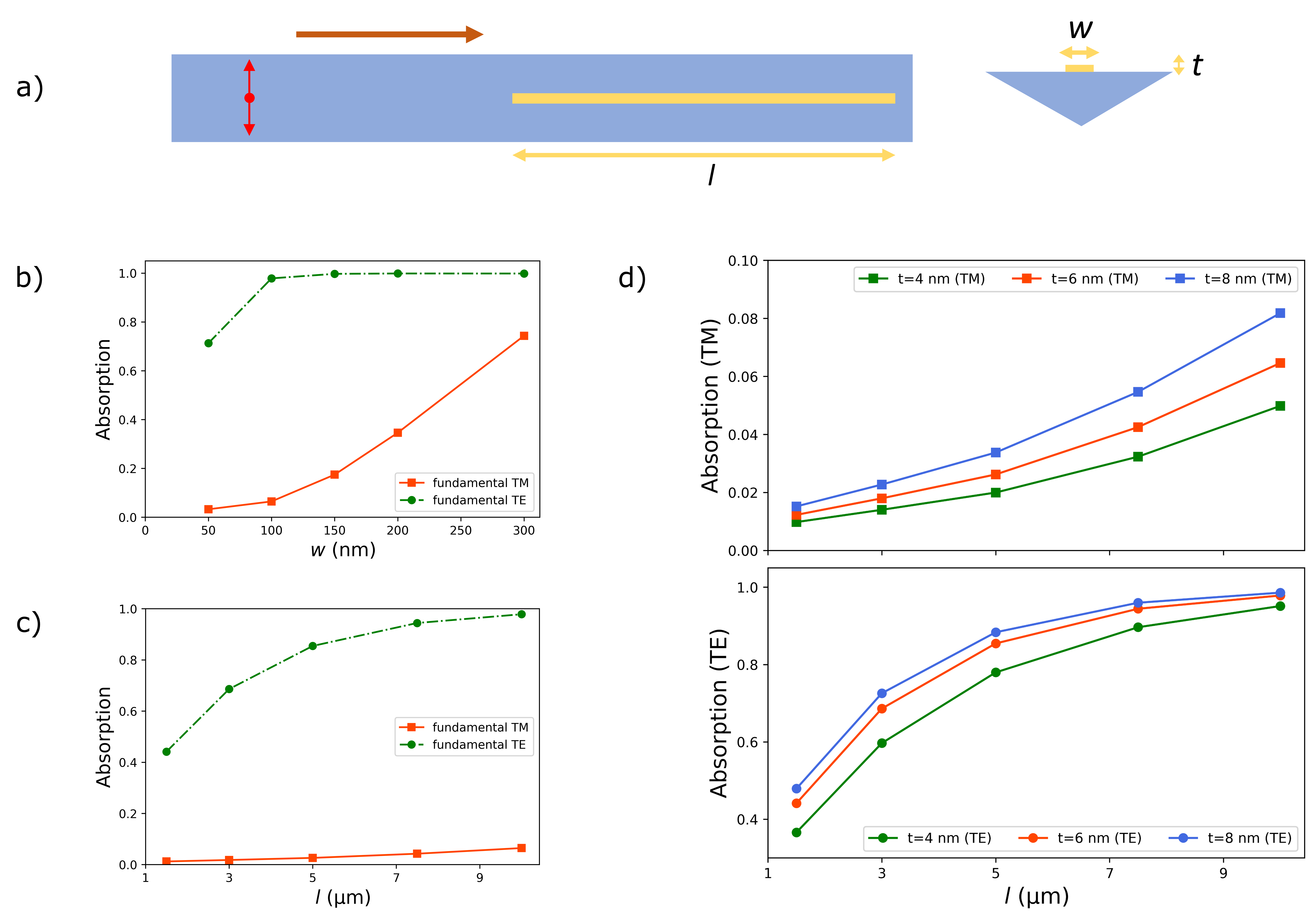}
    \caption{Simulated absorption of a single strip NbTiN layer. a) Top and cross-section view of a single strip NbTiN (yellow) with length $l$, width $w$ and thickness $t$, integrated on a triangular cross-section SiC waveguide (blue). Red dot indicates the color center and red arrows denotes its dipole orientation. Absorption of the fundamental TE and fundamental TM modes of the triangular waveguide as a function of b) width ($l$ = \SI{10}{\um}, $t$ = \SI{6}{\nm}). c) length ($w$ = \SI{100}{\nm}, $t$ = \SI{6}{\nm}). d) thickness ($w$ = \SI{100}{\nm}).
    }
    \label{fig:fig7}
\end{figure}

In this section, we look into the absorption efficiency of NbTiN based SNSPDs integrated on top of a triangular cross-section SiC waveguides based on the experimental measurements presented in Section \ref{SNSPD-exp}. 
This provides an upper limit for the achievable system detection efficiency of such an integrated detector. 
The absorption of color center emission (TE- or TM-like) for different configurations of NbTiN was modeled using the FDTD package in Lumerical software. 
The NbTiN was modeled using a complex permittivity model with ($n$, $k$) values obtained from the ellipsometry measurements discussed in the Section \ref{SNSPD-exp}. 
The top and cross-section view of a single strip of NbTiN integrated on top of a triangular cross-section waveguide are shown in Figure \ref{fig:fig7}a. 
The single strip has length $l$, width $w$, and thickness $t$, positioned at the center of the top surface of the waveguide. 

When the width of the NbTiN strip ($l$ = \SI{10}{\um}, $t$ = \SI{6}{\nm}) is increased, the absorption of the f-TE mode remains the same for widths greater than \SI{100}{\nm} and the absorption of the f-TM mode increases as the width increases, as shown in Figure \ref{fig:fig7}b. 
It should be noted that the absorption of the NbTiN strip is sensitive to the polarization of the mode in the waveguide. 
The evanescent fields of the f-TE mode are concentrated around the center of the top surface (inset of Figure \ref{fig:fig3}a), and therefore no changes in absorption are observed for larger SNSPD widths. 
In comparison, the evanescent fields of the f-TM mode are spread across the entire width of the waveguide (inset of Figure \ref{fig:fig3}c), requiring larger widths of NbTiN for efficient absorption. 
As the length ($w$ = \SI{100}{\nm}, $t$ = \SI{6}{\nm}) and thickness ($w$ = \SI{100}{\nm}) of the NbTiN strip increases, the absorption of both the f-TE and f-TM mode increases, as shown in Figure \ref{fig:fig7}c-d. 
Although larger thicknesses would result in better absorption, the single photon detection efficiency decreases \cite{Hofherr2010}, which is critical for applications in QIP. 

\begin{figure}[!ht]
    \centering
    \includegraphics[width = \textwidth]{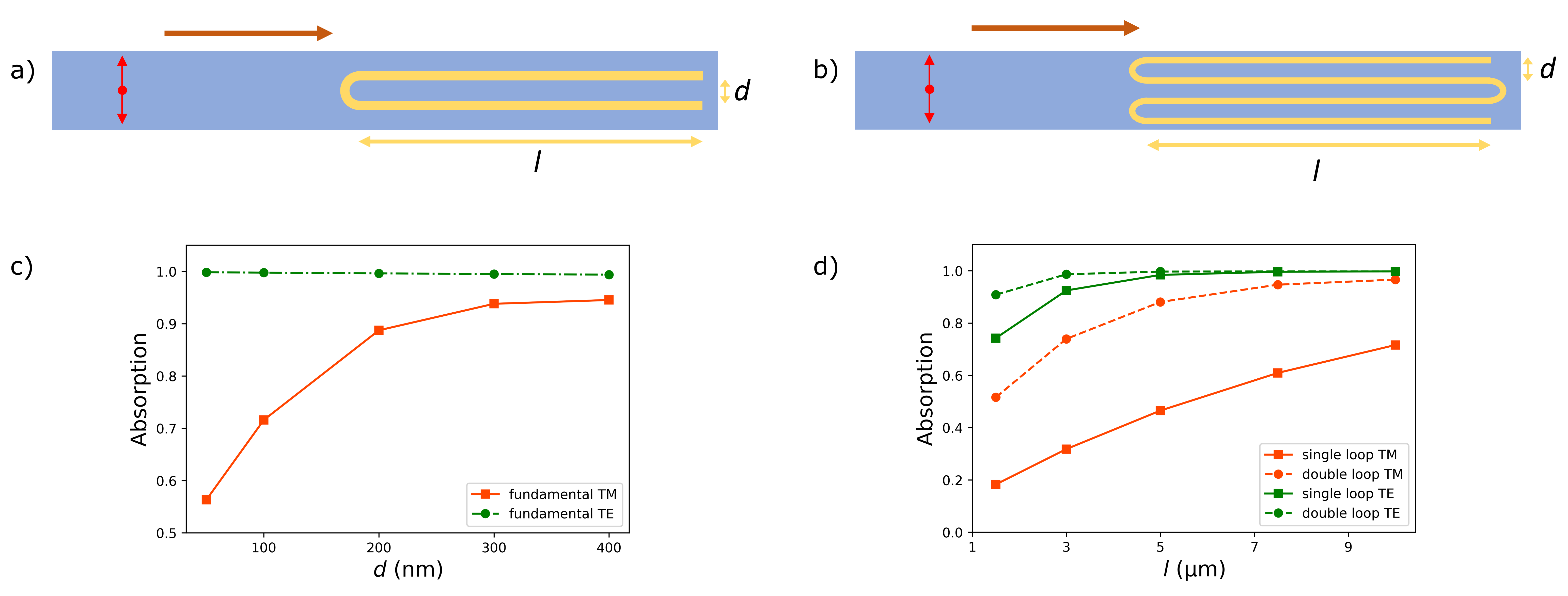}
    \caption{a) Top-view of a single loop NbTiN (yellow) with length $l$, spacing between adjacent arms $d$, on a triangular cross-section SiC waveguide (blue). b) Top-view of a double loop NbTiN (yellow) with length \textit{L}, spacing between two adjacent arms \textit{d}, on a triangular cross-section SiC waveguide (blue). Red dot indicates the color center and red arrows denotes its dipole orientation. c) Absorption of the fundamental TE and fundamental TM modes of the triangular waveguide as a function of spacing between adjacent arms, \textit{d} ($l$ = \SI{10}{\um}, $w$ = \SI{100}{\nm}, $t$ = \SI{6}{\nm}). d) Comparison of absorption of fundamental TE and fundamental TM modes of the triangular waveguide as a function of length, for single loop and double loop NbTiN ($w$ = \SI{100}{\nm}, $d$ = \SI{100}{\nm}, $t$ = \SI{6}{\nm}).
    }
    \label{fig:fig8}
\end{figure}

Typical SNSPD based detectors have nanowires fabricated into a meander shape to increase the optical detection area. 
To this end, we modeled single loop and double loop configurations of NbTiN integrated on top of a triangular cross-section SiC waveguides, as shown in Figures \ref{fig:fig8}a, \ref{fig:fig8}b, respectively. 
For a single loop NbTiN ($l$ = \SI{10}{\um}, $w$ = \SI{100}{\nm}, $t$ = \SI{6}{\nm}), as the spacing between two adjacent arms $d$ increases, the absorption of the f-TE mode remains constant and the absorption of the f-TM mode increases rapidly, as shown in Figure \ref{fig:fig8}c. 
The absorption of the double loop NbTiN ($w$ = \SI{100}{\nm}, $d$ = \SI{100}{\nm}, $t$ = \SI{6}{\nm}) is higher than the single loop NbTiN, with a significant improvement in the absorption of the f-TM mode, as shown in Figure \ref{fig:fig8}d. 
The polarization sensitivity of the double loop SNSPD at $l$ = \SI{10}{\um} almost disappears, thus having identical detection efficiencies for both f-TE and f-TM modes.

\section{Discussion}
The results presented in this paper demonstrate scalable approaches to efficiently collect and detect color center emission in triangular cross-section SiC devices for applications in QIP. 
Triangular cross-section devices are promising for chip scale integration of color center photonics. 
Moreover, these devices have a rich parameter space (etch angle, width) for optimal color center integration. 
At a given emission wavelength, there exists an optimal width for each triangular cross-section (with an etch angle alpha) such that the waveguide supports single mode propagation \cite{babin2022fabrication}, which is important for applications in QIP. 
The polarization of the color center emission depends on the dipole orientation of the color center with respect to the c-axis. 
Using 4H-SiC grown along a-plane, most of the color center emission (silicon vacancy, divacancy, nitrogen vacancy) can be coupled into the f-TE mode.

Furthermore, the coupling of the color center emission can be improved by positioning the color center at the maximum intensity point of the f-TE mode, which is around the centroid of the triangular cross-section.
Color centers are generated in SiC through implantation of high energy projectiles like electrons, protons, neutrons, ions. The implantation energy and the mass of the implanted particle determine the implantation damage done to the crystal, which can only be partially recovered through annealing \cite{gurfinkel2009ion}. 
Triangular cross-section devices can be designed depending on the implantation energies available, while still maintaining the single mode propagation and emitter's position at the centroid of the triangular cross-section. 
The waveguide simulated in this work ($w$ = \SI{800}{\nm}, $\alpha$ = \SI{60}{\degree}) has single mode propagation, with $\sim$\SI{44}{\percent} of the \SI{50}{\percent} dipole emission (\SI{1230}{\nm}) along one direction of the waveguide coupled to the f-TE mode. 

The collection efficiency ($\sim$\SI{44}{\percent}) on one end of the designed waveguide can be further improved using a photonic crystal mirror. 
The periodic variation of the dielectric constant (air holes) in a PC results in a photonic band gap for both TE- and TM-like modes in the waveguide. 
In a parallel work, a thorough photonic bandgap analysis using PWE method shows that the PCs can be used for a variety of applications and that their dimensions can be scaled depending on the color center emission wavelength \cite{saha2022photonic}. 
In triangular cross-section PCMs, the TE bandgap is an order of magnitude larger than the TM bandgap, consequently a relatively much smaller complete bandgap. 
Therefore, a more effective approach for using the PCs as reflectors of c-axis oriented color center emission is to fabricate them in a-plane SiC substrates \cite{babin2022fabrication}. 
This would ensure that a large portion of any color center emission is coupled to the f-TE mode, reflecting frequencies spanning the ZPL and PSB of the color center emission. 
FDTD analysis shows that the position of the emitter relative to the PCM determines whether there is constructive/destructive interference in the waveguide region. 
We highlight that using the PCM proposed in this paper, a randomly positioned emitter has a higher collection efficiency compared to a symmetric waveguide ($\sim$\SI{44}{\percent}) for \SI{75}{\percent} of cases.  

Efficient detection is another important factor for boosting the success rates of QIP protocols. 
SNSPDs offer a scalable approach for integrating detectors onto photonic devices. 
On-chip detection using SNSPDs can reduce optical losses, latency, and wiring complexity involved in off-chip detection \cite{najafi2015chip}. 
As a basis for \ce{SiC} waveguide integrated SNSPDs, we investigated, for the first time, the optical and electrical properties of \ce{NbTiN} thin films for SNSPD applications on 4H-\ce{SiC} substrates.
While the absorption of the \ce{NbTiN} films is significantly higher on \ce{SiC} than on \ce{SiO2} substrates (up to \SI{50}{\percent} for photons of \SI{1680}{\nm}), the magneto-electrical properties of the films, and thus presumably also of the detectors, are similar.
Provided that the fabrication of SNSPDs on \ce{SiC} waveguides is of similar quality as on \ce{SiO2} substrates, we expect that they provide comparable electrical performance and simultaneously enhanced system detection efficiency due to higher absorption in the film, making \ce{SiC} waveguide integrated SNSPDs a promising candidate for on-chip photon detection in QIP.

\section{Acknowledgements}
This work is supported by the National Science Foundation (CAREER-2047564), the Bavaria California Technology Center (BaCaTeC) - Internationalization of the High-Tech-Initiative, the German Federal Ministry of Education and Research via the funding program Quantum technologies - from basic research to market (contract numbers 16K1SQ033, 13N15855 and 13N15982) and via the project MARQUAND (contract number BN105022) and the Deutsche Forschungsgemeinschaft (DFG, German Research Foundation) under Germany’s Excellence Strategy – EXC-2111 – 390814868.

\section{Conflict of Interest}
The authors have no conflicts to disclose.

\section{Data Availability Statement}
The data that support the findings of this study are available upon reasonable request from the authors.

\end{document}